\documentclass[aps,preprint,prl,superscriptaddress]{revtex4}
\newif\ifpdf
\ifx\pdfoutput\undefined
	\pdffalse              %%normal LaTeX is executed
\else
	\pdfoutput=1           
	\pdftrue               %%pdfLaTeX is executed
\fi

\ifpdf
\else
\fi

\usepackage[T1]{fontenc}
\usepackage[latin1]{inputenc}
\ifpdf %%Inclusion of graphics via \includegraphics{file}
	\usepackage[pdftex]{graphicx} %%graphics in pdfLaTeX
\else
	\usepackage[dvips]{graphicx} %%graphics and normal LaTeX
\fi
\usepackage{amsmath}
\usepackage{amsthm}
\usepackage{amsfonts}

\begin{document}

%% File Extensions of Graphics %%%%%%%%%%%%%%%%%%%%%%%%%%%%%%
%% ==> This enables you to omit the file extension of a graphic.
%% ==> "\includegraphics{title.eps}" becomes "\includegraphics{title}".
%% ==> If you create 2 graphics with same content (but different file types)
%% ==> "title.eps" and "title.pdf", only the file processable by
%% ==> your compiler will be used.
%% ==> pdfLaTeX uses "title.pdf". LaTeX uses "title.eps".
%\ifpdf
%	\DeclareGraphicsExtensions{.pdf,.jpg,.png}
%\else
%	\DeclareGraphicsExtensions{.eps}
%\fi

%% Title Page %%%%%%%%%%%%%%%%%%%%%%%%%%%%%%%%%%%%%%%%%%%%%%%
%% ==> Write your text here or include other files.

%% The simple version:
\title{Impact of Phonon--Surface Roughness Scattering \\on Thermal Conductivity of Thin Si Nanowires}
\author{Pierre Martin}

\email[]{pmartin7@illinois.edu}
\affiliation{Beckman Institute for Advanced Technology and Science \\
Department of Electrical and Computer Engineering, %University of Illinois, Urbana-Champaign, Urbana, IL 61801, USA
}
\author{Zlatan Aksamija}
\affiliation{Beckman Institute for Advanced Technology and Science \\
Department of Electrical and Computer Engineering, %University of Illinois, Urbana-Champaign, Urbana, IL 61801, USA
}
\author{Eric Pop}
%\affiliation{Micro- and Nano-Technology Laboratory}
\email[]{epop@illinois.edu}
\affiliation{Beckman Institute for Advanced Technology and Science \\
Department of Electrical and Computer Engineering, %University of Illinois, Urbana-Champaign, Urbana, IL 61801, USA
}
\affiliation{Micro- and Nano-Technology Laboratory, \\
University of Illinois, Urbana-Champaign, Urbana, IL 61801, USA}
\author{Umberto Ravaioli}
\affiliation{Beckman Institute for Advanced Technology and Science \\
Department of Electrical and Computer Engineering, %University of Illinois, Urbana-Champaign, Urbana, IL 61801, USA
}
%\thanks{}


\date{\today}

%% Abstract Goes Here %%%%%%%%%%%%%%%%%%%%%%%%%%%%%%%%%%%%%%%
\begin{abstract}
We present a novel approach for computing the surface roughness-limited thermal conductivity of silicon nanowires with diameter $D$ < 100 nm. A frequency-dependent phonon scattering rate is computed from perturbation theory and related to a description of the surface through the root-mean-square roughness height $\Delta$ and autocovariance length $L$. Using a full-phonon dispersion relation, we find a quadratic dependence of thermal conductivity on diameter and roughness as $(D/\Delta)^2$. Computed results show excellent agreement with experimental data for a wide diameter and temperature range (25-350 K), and successfully predict the extraordinarily low thermal conductivity of 2 W m$^{-1}$K$^{-1}$ at room temperature in rough-etched 50 nm silicon nanowires.
\end{abstract}

\maketitle

%% Chapters %%%%%%%%%%%%%%%%%%%%%%%%%%%%%%%%%%%%%%%%%%%%%%%%%
%% ==> Write your text here or include other files.
Silicon nanowires (NW) have drawn much attention for their potential applications in field effect transistors \cite{ref1}, interconnects \cite{ref3}, thermoelectrics \cite{nature1, nature2, ref4, ref5, ChenPrl}, and heterostructures \cite{Dames}. Given their high surface-to-volume aspect ratio, the most prominent size effect limiting transport originates from electron or phonon surface scattering. In particular, being able to independently control electrical and thermal conductivity of such nanostructures through geometry, strain, or doping is extremely appealing for novel applications such as thermoelectrics and energy transport. Earlier data shows that reducing the NW diameter below 100 nm leads to a drastic reduction  in their electronic and thermal conductivity \cite{ref3, ref4, ChenPrl}. More puzzling are the recent experimental results of Hochbaum and Boukai \cite{nature1, nature2} which show that intentionally etched rough edges reduce the thermal conductivity of crystalline silicon NW  by a factor of about 100, to nearly the value of amorphous silicon. Several efforts have been previously made toward an accurate understanding of phonon-surface scattering \cite{ref5, ChenPrl, Dames}, however, no studies account for such experimental observations in very rough wires with diameter below 50 nm. Similarly, no model provides guidance on how the thermal conductivity of such NW scales with their surface roughness. 
	
	In this Letter we introduce a comprehensive approach to phonon-surface scattering in thin NW based on a perturbative treatment of interface roughness. We derive a matrix element for phonon-surface scattering which is directly related to a parametric description of surface roughness. Based on this approach, the effects of temperature change and surface quality on the phonon-surface scattering rate are evaluated in silicon NW of diameters below 100 nm. Resulting theoretical predictions of NW thermal conductivity shows excellent agreement with experimental values below 50 nm, where the effect of surface roughness is the strongest. 
	
	When the characteristic dimensions of asperities at a rough surface come to the order of the phonon or electron wavelength (5-30 \AA$ $), it is expected that the surface scattering rate will be altered to reflect the effect of the interface roughness. Such mechanisms are crucial to the understanding of electron transport in transistor inversion layers, where accurate models have been developed based on perturbation theory \cite{ref6,ref8}. While such formalism exists in the case of bulk phonon transport \cite{ref7}, current models of phonon-surface scattering in NW are based on simplified assumptions, most of them using the probability of diffuse scattering as a fitting parameter \cite{ref3, ref5}. Yet, the latter probability can be directly related to physical properties of the interface, which, among other options, may be experimentally observed by means of Transmission Electron Microscopy (TEM). Besides, it seems relevant that the effect of rough surfaces should be stronger in thin NW, and vary with the frequency of incident phonons.

%\begin{figure}[t!]
%	\centering
%		\includegraphics[width=0.65\columnwidth]{model.png}
%	\caption{Model of conduction in a Si nanowire with rough interfaces. Phonon transport is interrupted by scattering events originating from the %perturbations at the Si-SiO$_2$ interface. The roughness is described by its statistical root mean square $\Delta$ and length $L$.  }
%	\label{fig:model}
%\end{figure}

\begin{figure}[t]
 \begin{center}
  \includegraphics[width=0.65\columnwidth]{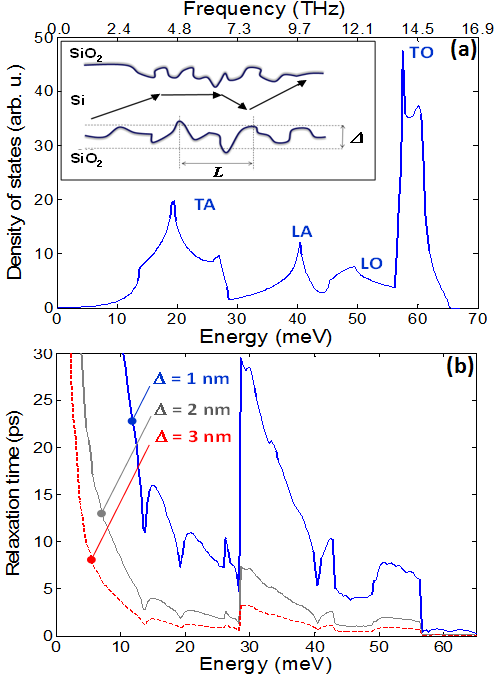}
  \end{center}
  \caption{a) Phonon density of states for silicon computed from a full phonon dispersion relation. b) Phonon relaxation time due to interface roughness for rms $\Delta=$ 1 nm, 2 nm, and 3 nm,  $T=$ 300 K, and a cross section equivalent to a circle of 115 nm diameter. $L$ is fixed to 6 nm.}
  \label{fig:scatrates}
\end{figure}

In a thin nanowire, variations of the confinement width perpendicular to the propagation direction influence phonon transport by perturbing the Hamiltonian of the system (Fig. \ref{fig:scatrates} (a) inset). It is assumed that boundary scattering is mainly an elastic process and no phonons are emitted into the surrounding environment. This condition reflects the case where NW are wrapped in a medium of considerably different thermal conductivity, as it is for Si NW in SiO$_2$ or vacuum. We model phonon transport in such Si NW of diameter below 115 nm, comparable to experimentally available data.  We introduce a new type of phonon scattering originating from the roughness of the NW surface. In essence, this scattering mechanism accounts for the fact that phonons ``see'' a rough NW as a series of constrictions along their propagation direction. In order to accurately model this effect at the nanometer scale, perturbation theory is used to derive the transition probability per unit time of an incident phonon of momentum $\mathbf{k}$ and energy $\hbar\omega$  to a new state of momentum $\mathbf{k'}$ and energy $\hbar\omega'$ due to the perturbed Hamiltonian $H'$ \cite{Ziman}:
\begin{equation}
P(\mathbf{k},\mathbf{k'}) = 2{|\langle \mathbf{k}|H'|\mathbf{k'}\rangle|}^2\frac{d}{dt}\left\{\frac{1-\cos[(\omega'-\omega)t/\hbar]}{(\omega'-\omega)^2}\right\}
\end{equation}  
Under a sufficiently long time $t$ in comparison to the energy relaxation time, the time derivative reduces to the Dirac delta function. The interface roughness is considered as a space varying dilation $\Delta(\mathbf{r})$ of the wire. This alters frequencies in a plane perpendicular to the propagation direction in such a way that $\omega(\mathbf{k}) = \omega_0(\mathbf{k})[1-\gamma\Delta]$  where $\gamma$ is a fitting constant determined from thermal expansion of the material, and $\omega_0(\mathbf{k})$ is the phonon dispersion of the unperturbed Hamiltonian. Following the derivation of Klemens \cite{ref7}, the matrix element for a perturbation due to a space varying dilation is
\begin{equation}
{|\langle \mathbf{k}|H'|\mathbf{k'}\rangle|}^2 = \frac{4 \gamma^2}{3V_{ol}}\omega'^2(\langle n'\rangle + 1)\Delta(\mathbf{\mathbf{k}-\mathbf{k'}})
\label{eq:nn}
\end{equation}   
where $V_{ol}$ is the volume of the device, and $\Delta(\mathbf{q})$ is the Fourier transform of the spatial perturbation, equal to $\Delta(\mathbf{q}) = \int{\Delta(\mathbf{r})e^{i\mathbf{q}.\mathbf{r}}}\,d\mathbf{r}$.
Additionally, the occupation number is given by the Bose-Einstein distribution and includes the temperature dependence of the scattering process $\langle n\rangle = \left(e^{\hbar\omega/k_BT}-1\right)^{-1}$.
	
	As shown by Goodnick et al. \cite{GoodnickFerry}, the autocovariance function of Si surface roughness is ``roughly'' fit by a Gaussian function, which, by the Wiener-Khinchin theorem yields a power spectrum of
\begin{equation}
\Delta(\mathbf{q}) = \pi\Delta^2L^2e^{-q^2L^2/4}
\label{eq:powerspec}
\end{equation}	 
where $\Delta$ is the root-mean-square (rms) value of the roughness fluctuations and $L$ is the autocovariance length, which is related to the mean distance between roughness peaks at the Si-SiO$_2$ interface (see Fig. \ref{fig:scatrates} (a) inset). In practice, these process-dependent parameters are experimentally set by the quality of the surface. Due to the $\omega'^2$ term in equation \ref{eq:nn}, low frequency phonons see little contribution from the surface perturbation. On the other hand, the power spectrum of equation \ref{eq:powerspec} favors scattering processes of the specular type. Hence, one can expect that there is a frequency range over which phonons experience a higher contribution from surface roughness scattering. The phonon scattering rate from a branch $i$ to a branch $j$ is given by
\begin{equation}
\tau_{i,j}^{-1}(E) = \int{P_{i,j}(\mathbf{k},\mathbf{k'})}\,d\mathbf{k'}
\end{equation}
	 
The volume integral over $\mathbf{k}'$-space can be reduced to a surface integral as shown in \cite{Kittel}:
\begin{equation}
\tau_{i,j}^{-1}(E) = \frac{2\pi}{\hbar N_i(E)}\int_{E'=E_i}{\frac{{|\langle\mathbf{k}|H'|\mathbf{k'}\rangle|}^2}{\nabla_{\mathbf{k'}}E'(\mathbf{k'})}}\,dS
\label{eq:scatrate}
\end{equation} 
where $N_i(E)$ is the phonon density of states in the $i_{th}$ branch, and $E'(\mathbf{k'})$ goes along the $j_{th}$ branch. The total scattering rate $\tau_i(E)$ starting in branch $i$ is the sum over all branches $j$ of the $\tau_{i,j}(E)$. 
	
	A Gilat-Raubenheimer (GR) scheme \cite{Gilat} is used to compute surface integrals, which constitutes the optimal trade-off between accuracy and computational efficiency. In order to carefully account for frequency dependence, a full phonon dispersion is used, which is obtained from an adiabatic bond charge model and tabulated for look-up \cite{Weber, Zlatan}. The GR method is also similarly applied to compute the phonon density of states based on the dispersion relation mentioned above. With this respect, the GR scheme divides the first Brillouin zone in a lattice of 40$\times$40$\times$40 cubes, which achieves sufficient accuracy in the 10-350 K range \cite{Zlatan}. Due to Bose-Einstein statistics, the accuracy of the scheme decays at lower temperatures, where finer grids or analytical derivation are required.
Modeled silicon NW have a square cross section with area equivalent to a circular cross section of diameter $D$, while the NW length is arbitrarily fixed to 2 $\mu$m. 
Transitions among all acoustic and optical branches are considered.

	The scattering rates are first computed for NW of equivalent diameter $D=$ 115 nm at $T=$ 300 K. For a fixed correlation length $L_0=$ 6 nm, phonon lifetime is calculated for increasing $\Delta$ in Fig. \ref{fig:scatrates}(b). Although no reliable data on the roughness of small wires is available yet, various theoretical and experimental studies have reported roughness rms ranging from 3 \AA$ $ to 5 nm in the case of extremely rough NW \cite{nature1, reggiani}. In this study we consider effective values for $\Delta$ in this range, and first use the 3 \AA$ $ value for ``smooth'' NW, while using an average of 3 nm for ``rough'' NW. Under the conditions cited above, the average phonon lifetime due to surface roughness alone is approximately 15 ps, and decreases with higher roughness rms values as $\Delta^{-2}$. As deduced from equation \ref{eq:powerspec}, a long correlation length $L$ favors scattering processes close to the specular type. While in the strong roughness limit where $\Delta/L>1$ the Gaussian approximation may be put at fault, the effect of the $L^2$ term tends to average out the contribution of $L$. We noticed only little deviation of the predicted thermal conductivity in the strong roughness limit, and consistently used a value of $L =$ 6 nm, which is estimated from the TEM images of Hochbaum et al. \cite{nature1} and provides a best fit in our case. 

	In our approach, an additional thermal variation of the phonon-surface scattering rate appears from Bose-Einstein statistics, where temperature delimits the occupation of each frequency range, thus retrieving the effect of occupation of lower energy branches at low temperature. Subsequently, it is possible to determine the thermal conductivity for NW of different cross sections \cite{Glassbrener}. The contribution to thermal conductivity of branch $i$ is
\begin{equation}
\kappa_i(T) = \frac{1}{3}\int_{i}{EN(E)\frac{d\langle n \rangle}{dT}v_i(E)\tau^{tot}_i(E)}\,dE
\end{equation}	
where $v_i$ is the velocity of sound which is dependent on the direction of propagation, here assumed to be in the <001> direction. In order to reproduce the measured physical behavior of the NW in the 10-350 K temperature range, Umklapp, normal, impurity, boundary, and surface roughness scattering mechanisms have been considered in the derivation of the branch-specific scattering time, as summarized in Table \ref{tab:scat}. Umklapp scattering in transverse acoustic branches is efficiently described with the law derived in Ref. \cite{Mingo}, which has shown good agreement with Si NW experiments. Additionally, in the temperature range considered, normal scattering in longitudinal acoustic branches is accounted for according to the derivation of Holland \cite{Holland}. A $C\omega^4$ law is often used in the literature for impurity scattering, and we found that a constant $C=$ 8$\times10^{-45}$ s/K$^4$ fits experimental data with more accuracy than the value of 1.05$\times10^{-44}$ s/K$^4$ resulting from the derivation of Ref \cite{Holland}. In the case of NW etched  from a 10 $\Omega$-cm wafer \cite{nature1}, the low doping concentration has almost no impact on the value of $C$. Besides, the boundary scattering rate depends on the sound velocity in a given branch, and is consequently frequency dependent. An analytical expression is used for the sound velocity in acoustic branches, which is derived from \cite{Pop} for its good fit with bulk Si data in the <001> direction.

\begin{figure}[t]
 \begin{center}
  \includegraphics[width=0.65\columnwidth]{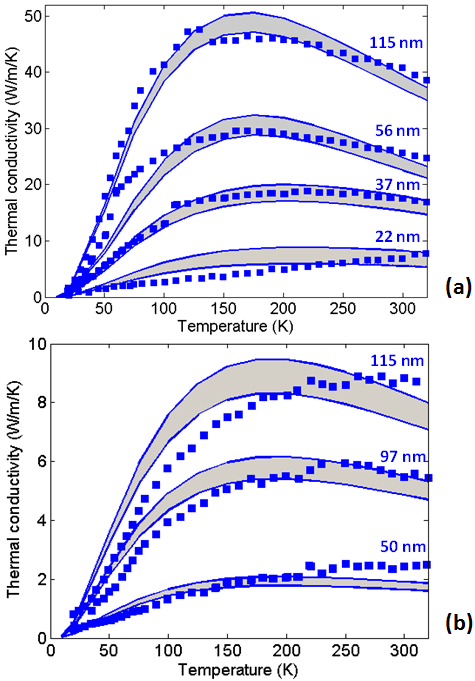}
 \end{center}
  \caption
  %{a) Thermal conductivity of smooth vapor-liquid solid silicon nanowires ($\Delta=$ 3 \AA$ $, $L=$ 6nm). Solid lines are theoretical predictions %including a surface roughness scattering term, blue squares are taken from \cite{ref4}. b) Thermal conductivity of rough EE silicon nanowires %($\Delta=$ 3 nm, $L=$ 6 nm). Squares are taken from \cite{nature1}. Simulation and experimental data are compared at similar cross sections.}
{(a) Thermal conductivity of smooth VLS Si NW. Shaded areas are theoretical predictions with roughness r.m.s.  $\Delta =$ 1 - 3 \AA$ $, blue squares are taken from \cite{ref4}. (b) Thermal conductivity of rough EE Si NW ($\Delta  =$ 3 - 3.25 nm). Squares are taken from \cite{nature1}. Simulation and experimental data are compared at similar cross sections. $L =$ 6 nm.}  
\label{fig:fig2}
\end{figure}

	Fig. \ref{fig:fig2} (a) and \ref{fig:fig2} (b) compare thermal conductivity computed from the model presented above vs. experimental data from \cite{ref4} and \cite{nature1}, for NW ranging from 115 nm to 22 nm diameter. As nothing guarantees that all NW have similar roughness parameters, we considered smooth NW with 1 \AA$ $ $<\Delta<$ 3 \AA, and rough NW with 3 nm $<\Delta<$ 3.25 nm.  A good fit is found for smooth NW grown by vapor-liquid-solid (VLS) mechanism with diameter above 37 nm. Similarly, the model reproduces the drastic decrease in thermal conductivity for rough electroless-etching (EE) NW presented in \cite{nature1}. Higher discrepancy is found for the 22 nm smooth VLS NW, for which the sensitivity to surface roughness is expected to be higher. Besides, the perturbative approach remains valid as long as perturbations remain small in comparison to the total phonon energy. For low temperature phonons and nanowires of diameter below 20 nm, explicit quantum treatment may be required. Since surface roughness scattering has little impact on low energy phonons, additional low temperature discrepancy is attributed to impurity and classical boundary scattering.

\begin{figure}[t]
 \begin{center}
  \includegraphics[width=0.99\columnwidth]{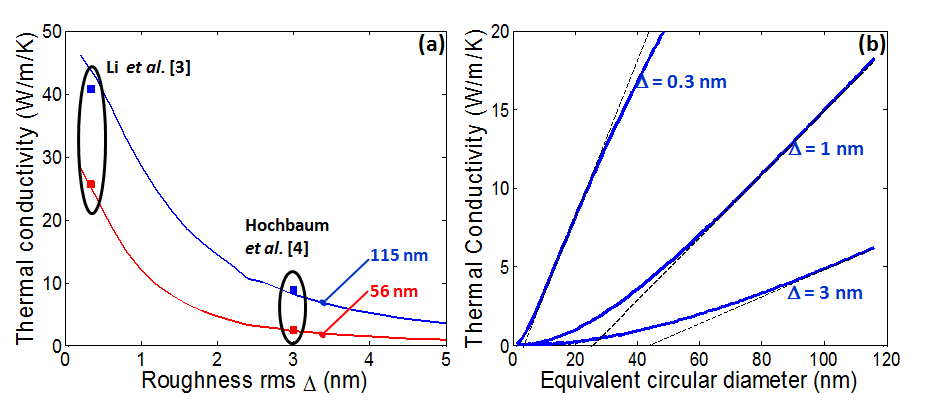}
 \end{center}
  \caption{a) Predicted effect of roughness rms on the thermal conductivity of 115 nm and 56 nm NW at $T=$ 300 K, $L =$ 6 nm. Simulation and experimental data are compared at similar cross sections. (b)  Effect of NW equivalent circular diameter on thermal conductivity ($L =$ 6 nm).}
\label{fig:fig3}
\end{figure}

\begin{figure}[t]
	\centering
		\includegraphics[width=0.6\columnwidth]{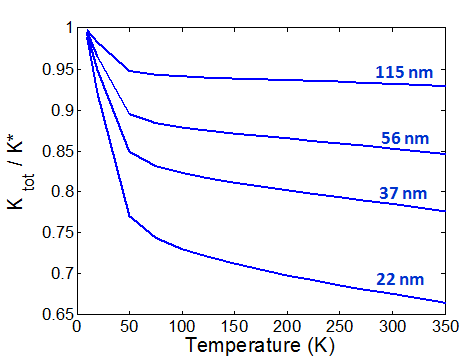}
	\caption{ Proportional contribution of the surface roughness scattering term to the thermal conductivity of VLS NW ($\Delta=$ 3 \AA$ $, $L=$ 6 nm). The reference value $K^*$ is the thermal conductivity of a perfectly smooth wire ($\Delta=$ 0 nm).}
	\label{fig:InfluenceRoughness}
\end{figure}

	It is important to point out that our approach based on perturbation theory introduces a dependence in $(D/\Delta)^2$ of the NW thermal conductivity, in contrast to the typical linear scaling with $D$ used in previous descriptions. In this scope, figures \ref{fig:fig3} (a) and \ref{fig:fig3} (b) show how the thermal conductivity is lowered by concurrent effects of small diameter and rough surfaces. In particular, there exists a critical diameter below which the roughness-limited thermal conductivity of the NW noticeably deviates from the classical linear approximation. The fact that this critical diameter increases with higher roughness rms $\Delta$, in turn, supports the expectation that heat conduction at small NW scales is strongly limited by their surface roughness. This assumption is further justified in Fig. \ref{fig:InfluenceRoughness}, where it is observed that the total contribution of surface roughness to limiting the thermal conductivity is increased about 7 times from the 115 nm to the 22 nm case. Finally, the temperature dependence that arises from the roughness scattering rate is also retrieved. Thus, at low temperatures, low frequency phonons see little effect from the perturbations at the silicon interface.

In summary, we have demonstrated a remarkably strong effect of surface roughness on the thermal conductivity of thin silicon nanowires. Based on a full phonon dispersion relation we introduced a novel frequency-dependent model of boundary scattering for phonons. The resulting simulated thermal conductivity of NW of diameter $D<$ 115 nm shows excellent agreement with recent experimental work. In particular, at low NW diameters, we predict a strong deviation of the roughness-limited thermal conductivity from the linear diameter dependence ($\sim D$) to a scaling as $ (D/\Delta)^2$, where $\Delta$ is the rms surface roughness. The approach presented here can be generally extended to NW of various materials, sizes, and direction of thermal propagation using the same full-phonon dispersion relation with the inclusion of the surface-scattering rate in a Monte Carlo simulation.  

\begin{acknowledgments}
This work was supported in part by the DARPA through the IMPACT Center for Advancement
of MEMS/NEMS VLSI grant n. HR0011-06-1-0046 (PM), the DOE
Computational Science Graduate Fellowship Program
of the OSNNSA under
contract DE-FG02-97ER25308 (ZA), and the Nanoelectronics Research Initiative (NRI) SWAN center (EP).
\end{acknowledgments}

%%%%%%%%%%%%%%%%%%%%%%%%%%%%%%%%%%%%%%%%%%%%%%%%%%%%%%%%%%%%%
%% BIBLIOGRAPHY AND OTHER LISTS
%%%%%%%%%%%%%%%%%%%%%%%%%%%%%%%%%%%%%%%%%%%%%%%%%%%%%%%%%%%%%
%% A small distance to the other stuff in the table of contents (toc)
%\addtocontents{toc}{\protect\vspace*{\baselineskip}}

%% The Bibliography
%% ==> You need a file 'literature.bib' for this.
%% ==> You need to run BibTeX for this (Project | Properties... | Uses BibTeX)
%\addcontentsline{toc}{chapter}{Bibliography} %'Bibliography' into toc
%\nocite{*} %Even non-cited BibTeX-Entries will be shown.
%\bibliographystyle{alpha} %Style of Bibliography: plain / apalike / amsalpha / ...
%\bibliography{literature} %You need a file 'literature.bib' for this.

%% The List of Figures
%\clearpage
%\addcontentsline{toc}{chapter}{List of Figures}
%\listoffigures

%% The List of Tables
%\clearpage
%\addcontentsline{toc}{chapter}{List of Tables}
%\listoftables

%%%%%%%%%%%%%%%%%%%%%%%%%%%%%%%%%%%%%%%%%%%%%%%%%%%%%%%%%%%%%
%% APPENDICES
%%%%%%%%%%%%%%%%%%%%%%%%%%%%%%%%%%%%%%%%%%%%%%%%%%%%%%%%%%%%%
%\appendix
%% ==> Write your text here or include other files.

%\input{FileName} %You need a file 'FileName.tex' for this.

\bibliography{biblio}

\begin{thebibliography}{22}
\expandafter\ifx\csname natexlab\endcsname\relax\def\natexlab#1{#1}\fi
\expandafter\ifx\csname bibnamefont\endcsname\relax
  \def\bibnamefont#1{#1}\fi
\expandafter\ifx\csname bibfnamefont\endcsname\relax
  \def\bibfnamefont#1{#1}\fi
\expandafter\ifx\csname citenamefont\endcsname\relax
  \def\citenamefont#1{#1}\fi
\expandafter\ifx\csname url\endcsname\relax
  \def\url#1{\texttt{#1}}\fi
\expandafter\ifx\csname urlprefix\endcsname\relax\def\urlprefix{URL }\fi
\providecommand{\bibinfo}[2]{#2}
\providecommand{\eprint}[2][]{\url{#2}}

\bibitem[{\citenamefont{Singhan et~al.}(2006)\citenamefont{Singhan, Agarwal,
  Bera et~al.}}]{ref1}
\bibinfo{author}{\bibfnamefont{N.}~\bibnamefont{Singhan}},
  \bibinfo{author}{\bibfnamefont{A.}~\bibnamefont{Agarwal}},
  \bibinfo{author}{\bibfnamefont{L.}~\bibnamefont{Bera}}, \bibnamefont{et~al.},
  \bibinfo{journal}{IEEE Electron Dev. Lett.} \textbf{\bibinfo{volume}{27}},
  \bibinfo{pages}{383} (\bibinfo{year}{2006}).

\bibitem[{\citenamefont{Steinghol}(2005)}]{ref3}
\bibinfo{author}{\bibfnamefont{W.}~\bibnamefont{Steinghol}},
  \bibinfo{journal}{Appl. Phys. Lett.} \textbf{\bibinfo{volume}{97}}
  (\bibinfo{year}{2005}).

\bibitem[{\citenamefont{Hochbaum et~al.}(2008)\citenamefont{Hochbaum, Chen,
  Delgado et~al.}}]{nature1}
\bibinfo{author}{\bibfnamefont{A.~I.} \bibnamefont{Hochbaum}},
  \bibinfo{author}{\bibfnamefont{R.}~\bibnamefont{Chen}},
  \bibinfo{author}{\bibfnamefont{R.}~\bibnamefont{Delgado}},
  \bibnamefont{et~al.}, \bibinfo{journal}{Nature}
  \textbf{\bibinfo{volume}{451}}, \bibinfo{pages}{163} (\bibinfo{year}{2008}).

\bibitem[{\citenamefont{Boukai et~al.}(2008)\citenamefont{Boukai, Bunimovich,
  Tahir-Kheli et~al.}}]{nature2}
\bibinfo{author}{\bibfnamefont{A.}~\bibnamefont{Boukai}},
  \bibinfo{author}{\bibfnamefont{Y.}~\bibnamefont{Bunimovich}},
  \bibinfo{author}{\bibfnamefont{J.}~\bibnamefont{Tahir-Kheli}},
  \bibnamefont{et~al.}, \bibinfo{journal}{Nature}
  \textbf{\bibinfo{volume}{451}}, \bibinfo{pages}{168} (\bibinfo{year}{2008}).

\bibitem[{\citenamefont{Li et~al.}(2003)}]{ref4}
\bibinfo{author}{\bibfnamefont{D.}~\bibnamefont{Li}} \bibnamefont{et~al.},
  \bibinfo{journal}{Appl. Phys. Lett.} \textbf{\bibinfo{volume}{83}},
  \bibinfo{pages}{2934} (\bibinfo{year}{2003}).

\bibitem[{\citenamefont{Lacroix et~al.}(2004)}]{ref5}
\bibinfo{author}{\bibfnamefont{D.}~\bibnamefont{Lacroix}} \bibnamefont{et~al.},
  \bibinfo{journal}{Appl. Phys. Lett.} \textbf{\bibinfo{volume}{89}}
  (\bibinfo{year}{2004}).

\bibitem[{\citenamefont{Chen et~al.}(2008)\citenamefont{Chen, Hochbaum, Murphy
  et~al.}}]{ChenPrl}
\bibinfo{author}{\bibfnamefont{R.}~\bibnamefont{Chen}},
  \bibinfo{author}{\bibfnamefont{A.}~\bibnamefont{Hochbaum}},
  \bibinfo{author}{\bibfnamefont{P.}~\bibnamefont{Murphy}},
  \bibnamefont{et~al.}, \bibinfo{journal}{Phys. Rev. Lett.}
  \textbf{\bibinfo{volume}{101}} (\bibinfo{year}{2008}).

\bibitem[{\citenamefont{Dames and Chen}(2004)}]{Dames}
\bibinfo{author}{\bibfnamefont{C.}~\bibnamefont{Dames}} \bibnamefont{and}
  \bibinfo{author}{\bibfnamefont{G.}~\bibnamefont{Chen}}, \bibinfo{journal}{J.
  Appl. Phys.} \textbf{\bibinfo{volume}{95}}, \bibinfo{pages}{682}
  (\bibinfo{year}{2004}).

\bibitem[{\citenamefont{Goodnick and Ferry}(1997)}]{ref6}
\bibinfo{author}{\bibfnamefont{S.}~\bibnamefont{Goodnick}} \bibnamefont{and}
  \bibinfo{author}{\bibfnamefont{D.}~\bibnamefont{Ferry}},
  \emph{\bibinfo{title}{Transport in Nanostructures}}
  (\bibinfo{publisher}{Cambridge University Press}, \bibinfo{year}{1997}),
  \bibinfo{edition}{1st} ed.

\bibitem[{\citenamefont{Yamakawa et~al.}(1996)}]{ref8}
\bibinfo{author}{\bibfnamefont{S.}~\bibnamefont{Yamakawa}}
  \bibnamefont{et~al.}, \bibinfo{journal}{Journ. Appl. Phys.}
  \textbf{\bibinfo{volume}{79}} (\bibinfo{year}{1996}).

\bibitem[{\citenamefont{Klemens}(1957)}]{ref7}
\bibinfo{author}{\bibfnamefont{P.}~\bibnamefont{Klemens}},
  \bibinfo{journal}{Solid State Physics} \textbf{\bibinfo{volume}{7}},
  \bibinfo{pages}{1} (\bibinfo{year}{1957}).

\bibitem[{\citenamefont{Ziman}(1960)}]{Ziman}
\bibinfo{author}{\bibfnamefont{J.}~\bibnamefont{Ziman}},
  \emph{\bibinfo{title}{Electrons and phonons: the theory of transport
  phenomena in solids}} (\bibinfo{publisher}{Clarendon}, \bibinfo{year}{1960}),
  \bibinfo{edition}{1st} ed.

\bibitem[{\citenamefont{Goodnick et~al.}(1983)}]{GoodnickFerry}
\bibinfo{author}{\bibfnamefont{S.}~\bibnamefont{Goodnick}}
  \bibnamefont{et~al.}, \bibinfo{journal}{Journ. Vacuum Sci. Tech. B}
  \textbf{\bibinfo{volume}{1}}, \bibinfo{pages}{803} (\bibinfo{year}{1983}).

\bibitem[{\citenamefont{Kittel}(2005)}]{Kittel}
\bibinfo{author}{\bibfnamefont{C.}~\bibnamefont{Kittel}},
  \emph{\bibinfo{title}{Introduction to solid state physics}}
  (\bibinfo{publisher}{Wiley}, \bibinfo{year}{2005}), \bibinfo{edition}{8th}
  ed.

\bibitem[{\citenamefont{Gilat and Raubenheimer}(1966)}]{Gilat}
\bibinfo{author}{\bibfnamefont{G.}~\bibnamefont{Gilat}} \bibnamefont{and}
  \bibinfo{author}{\bibfnamefont{L.}~\bibnamefont{Raubenheimer}},
  \bibinfo{journal}{Phys. Rev.} \textbf{\bibinfo{volume}{144}}
  (\bibinfo{year}{1966}).

\bibitem[{\citenamefont{Weber}(1970)}]{Weber}
\bibinfo{author}{\bibfnamefont{W.}~\bibnamefont{Weber}},
  \bibinfo{journal}{Phys. Rev. B} \textbf{\bibinfo{volume}{15}}
  (\bibinfo{year}{1970}).

\bibitem[{\citenamefont{Aksamija and Ravaioli}(2007)}]{Zlatan}
\bibinfo{author}{\bibfnamefont{Z.}~\bibnamefont{Aksamija}} \bibnamefont{and}
  \bibinfo{author}{\bibfnamefont{U.}~\bibnamefont{Ravaioli}},
  \bibinfo{journal}{Electro/Information Tech.} \textbf{\bibinfo{volume}{1}},
  \bibinfo{pages}{70} (\bibinfo{year}{2007}).

\bibitem[{\citenamefont{Lenzi et~al.}(2008)\citenamefont{Lenzi, Palestri, Gnani
  et~al.}}]{reggiani}
\bibinfo{author}{\bibfnamefont{M.}~\bibnamefont{Lenzi}},
  \bibinfo{author}{\bibfnamefont{P.}~\bibnamefont{Palestri}},
  \bibinfo{author}{\bibfnamefont{E.}~\bibnamefont{Gnani}},
  \bibnamefont{et~al.}, \bibinfo{journal}{IEEE Trans. Electron. Dev.}
  \textbf{\bibinfo{volume}{55}}, \bibinfo{pages}{2086} (\bibinfo{year}{2008}).

\bibitem[{\citenamefont{Glassbrener and Slack}(1964)}]{Glassbrener}
\bibinfo{author}{\bibfnamefont{C.}~\bibnamefont{Glassbrener}} \bibnamefont{and}
  \bibinfo{author}{\bibfnamefont{A.}~\bibnamefont{Slack}},
  \bibinfo{journal}{Phys. Rev.} \textbf{\bibinfo{volume}{134}},
  \bibinfo{pages}{1058} (\bibinfo{year}{1964}).

\bibitem[{\citenamefont{Mingo}(2003)}]{Mingo}
\bibinfo{author}{\bibfnamefont{N.}~\bibnamefont{Mingo}},
  \bibinfo{journal}{Phys. Rev. B} \textbf{\bibinfo{volume}{68}},
  \bibinfo{pages}{168} (\bibinfo{year}{2003}).

\bibitem[{\citenamefont{Holland}(1963)}]{Holland}
\bibinfo{author}{\bibfnamefont{M.}~\bibnamefont{Holland}},
  \bibinfo{journal}{Phys. Rev} \textbf{\bibinfo{volume}{132}},
  \bibinfo{pages}{2461} (\bibinfo{year}{1963}).

\bibitem[{\citenamefont{Pop et~al.}(2004)\citenamefont{Pop, Dutton, and
  Goodson}}]{Pop}
\bibinfo{author}{\bibfnamefont{E.}~\bibnamefont{Pop}},
  \bibinfo{author}{\bibfnamefont{R.}~\bibnamefont{Dutton}}, \bibnamefont{and}
  \bibinfo{author}{\bibfnamefont{K.}~\bibnamefont{Goodson}},
  \bibinfo{journal}{Journ. App. Phys.} \textbf{\bibinfo{volume}{96}},
  \bibinfo{pages}{4998} (\bibinfo{year}{2004}).

\end{thebibliography}

\newpage

\begin{table}
\begin{ruledtabular}
		\begin{tabular}{| l|c r |}
			\hline
			\small{Mechanism} & \small{Analytical Model}  & \small{Constants} \\
			\hline
			\small{Phonon-Phonon:} &  &\\ 
			 \small{\emph{Longitudinal} \cite{Holland}}	& $B_LT^3\omega^2$  & $B_L$=2$\times 10^{-24}$ s/K$^3$\\
			 \small{\emph{Transverse} \cite{Mingo}} & $B_TTe^{-\theta_1/T}\omega^2$  & $B_T$=1.73$\times 10^{-19}$ s/K\\
			  &  & $\theta_1$ = 137.3 K \\
			%&  & \\
			\small{Impurity \cite{Holland}} & $C\omega^4$  & $C$=8$\times 10^{-45}$ s/K$^{4}$\\
			%& &  \\
			\small{Boundary} & $v_iD^{-1}\sqrt{1+4\omega c_i/v_i}$  & $v_L$=9.01$\times 10^3$ m/s\\
			 \small{\cite{Glassbrener, Pop}}&  & $v_T$=5.23$\times 10^3$ m/s\\
			 &  & $c_L$=-2$\times10^{-7}$ m/s$^{2}$\\
			 &  & $c_T$=-2.26$\times 10^{-7}$ m/s$^{2}$\\
			%&  & \\
			\small{Surf. Roughness} & cf. Eq. \ref{eq:scatrate} & $\Delta$= 3 - 50 \AA\\
			 &  & $L$ = 60 \AA\\
			\hline
%		\end{tabular}
	\end{tabular}
	\end{ruledtabular}
	\caption{Summary of the scattering processes considered in the derivation of the thermal conductivity.}
	\label{tab:scat}
\end{table}


\end{document}